\documentclass[12pt,preprint]{aastex}
	\usepackage{natbib}

	\renewcommand{\deg}{\mbox{$^{\circ}$}}

	\def\deg{\ifmmode^\circ\else$^\circ$\fi}    %Degree sign%
	
	\def\hper{\ifmmode \rlap.^{h}\else $\rlap{.}^h$\fi} 
	\def\sper{\ifmmode \rlap.^{s}\else $\rlap{.}^s$\fi}    %Superscript 's' over per
	\def\deg{${}^\circ$}

	\def\bmv{\hbox{\it B--V\/}}

	\def\today{\number\year\space \ifcase\month\or
	  January\or February\or March\or April\or May\or June\or
	  July\or August\or September\or October\or November\or December\fi
	  \space\number\day}
	\def\now{\number\year\space \ifcase\month\or
	  January\or February\or March\or April\or May\or June\or
	  July\or August\or September\or October\or November\or December\fi
	  \space\number\day .\number\time}

	\shorttitle{On the $\Delta V_{HB}^{bump}$ parameter in GCs} 
	\shortauthors{Di Cecco et al.}

\begin{document}
	\title{On the $\Delta V_{HB}^{bump}$ parameter in Globular Clusters\altaffilmark{1}}

	\author{
	A.\ Di Cecco\altaffilmark{2,3},
	G.\ Bono\altaffilmark{2,3},
	P.\ B.\ Stetson\altaffilmark{4},
	A.\ Pietrinferni\altaffilmark{5},
	R.\ Becucci\altaffilmark{6},
	S.\ Cassisi\altaffilmark{5},
	S.\ Degl'Innocenti\altaffilmark{6,7}, 
	G.\ Iannicola\altaffilmark{3},
	P.\ G.\ Prada Moroni\altaffilmark{6,7},
	R.\ Buonanno\altaffilmark{2,8},
	A.\ Calamida\altaffilmark{9},
	F.\ Caputo\altaffilmark{3},
	M.\ Castellani\altaffilmark{3},
	C.\ E.\ Corsi\altaffilmark{3},
	I.\ Ferraro\altaffilmark{3},
	M.\ Dall'Ora\altaffilmark{10},
	M.\ Monelli\altaffilmark{11},
	M.\ Nonino\altaffilmark{12},
	A.\ M.\ Piersimoni\altaffilmark{5},
	L.\ Pulone\altaffilmark{3},
	M.\ Romaniello\altaffilmark{9},
	M.\ Salaris\altaffilmark{13},
	A.\ R.\ Walker\altaffilmark{14}, and
	M.\ Zoccali\altaffilmark{15}
	}

	\altaffiltext{1}{Based in part on data obtained from the ESO/ST-ECF Science 
	Archive Facility, from the Isaac Newton Group Archive which is maintained as part 
	of the CASU Astronomical Data Centre at the Institute of Astronomy, Cambridge
	and from the Canadian Astronomy Data Centre operated by the National Research 
	Council of Canada with the support of the Canadian Space Agency.}
	\altaffiltext{2}{Dipartimento di Fisica, Universit\`a di Roma Tor Vergata, via della 
	Ricerca Scientifica 1, 00133 Rome, Italy; alessandra.dicecco@roma2.infn.it}
	\altaffiltext{3}{INAF--OAR, via Frascati 33, Monte Porzio Catone, Rome, Italy}
	\altaffiltext{4}{DAO--HIA, NRC, 5071 West Saanich Road, Victoria, BC V9E 2E7, Canada}
	\altaffiltext{5}{INAF--OACTe, via M. Maggini, 64100 Teramo, Italy}
	\altaffiltext{6}{Dipartimento di Fisica, Universit\`a di Pisa, Largo B. Pontecorvo 2, 56127 Pisa, Italy}
	\altaffiltext{7}{INFN--Pisa, via E. Fermi 2, 56127 Pisa, Italy}
	\altaffiltext{8}{ASI--Science Data Center, ASDC c/o ESRIN, via G. Galilei, 00044 Frascati, Italy}
	\altaffiltext{9}{ESO, Karl-Schwarzschild-Str. 2, 85748 Garching bei Munchen, Germany}
	\altaffiltext{10}{INAF--OACN, via Moiariello 16, 80131 Napoli, Italy}
	\altaffiltext{11}{IAC, Calle Via Lactea, E38200 La Laguna, Tenerife, Spain}
	\altaffiltext{12}{INAF--OAT, via G.B. Tiepolo 11, 40131 Trieste, Italy}
	\altaffiltext{13}{ARI, Liverpool John Moores University, Twelve Quays House, Birkenhead CH41 1LD}
	\altaffiltext{14}{CTIO--NOAO, Casilla 603, La Serena, Chile}
	\altaffiltext{15}{PUC, Departamento de Astronomia y Astrofisica, Casilla 306, Santiago 22, Chile}

	\date{\centering drafted \today\ / Received / Accepted }

	\begin{abstract}
	We present new empirical estimates of the $\Delta V_{HB}^{bump}$ parameter for
	15 Galactic globular clusters (GGCs) using accurate and homogeneous ground-based
	optical data. Together with similar evaluations available in the literature, we
	ended up with a sample of 62 GGCs covering a very broad range in metal content
	(--2.16$\le$[M/H]$\le$--0.58 dex). Adopting the homogeneous metallicity scale
	provided either by Kraft \& Ivans (2004) or by 
	% Point B 
	Carretta et al.\ (2009), we found that the observed $\Delta V_{HB}^{bump}$ 
	parameters are larger than predicted.  In the metal-poor regime 
	([M/H]$\lesssim$--1.7, --1.6 dex) 40\% of GCs show discrepancies 
	of $2\sigma$ ($\approx$0.40 mag) or more. Evolutionary models that 
	account either for $\alpha$- and CNO-enhancement or for helium 
	enhancement do not alleviate the discrepancy between theory and observations.
	The outcome is the same if we use the new Solar heavy-element mixture. 
	The comparison between $\alpha$- and CNO-enhanced evolutionary 
	models and observations in the Carretta et al.\ metallicity scale also 
	indicates that observed $\Delta V_{HB}^{bump}$ parameters, in the metal-rich 
	regime ([M/H]$\ge$0), might be systematically smaller than predicted.   
	\end{abstract}

	\keywords{globular clusters: general --- stars: evolution --- stars: 
	horizontal-branch --- stars: Population II}

	\maketitle

	%%%%%%%%%%%%%%%%%%%%%%%%%%%%%%%%%%%%%%%%%%%%%%%%%%%%%%%%%%%%%%%%%%%%%
	\section{Introduction}
	Globular clusters (GCs) possess several evolutionary 
	features that show up in the color-magnitude diagram (CMD) 
	and/or in the luminosity function (LF). Along  
	the red giant branch (RGB) there is a well defined 
	bump in the differential LF and, equivalently, a change in the slope 
	of the cumulative LF. This occurs when the   
	H-burning shell crosses the chemical discontinuity left behind 
	by the deepest penetration of the convective envelope (first 
	dredge-up) at the base of the RGB (Thomas 1967; 
	Iben 1968; Renzini  \&  Fusi Pecci 1988; Castellani et al.\ 1989). 
	The efficiency of the H-burning shell is affected by the sharp 
	increase in the hydrogen abundance and the stellar luminosity 
	experiences a temporary drop. 
	Stars thus cross the same luminosity range three times, and 
	a bump/overdensity appears in the differential LF of 
	RGB stars. 

	The RGB bump is interesting because:  
	{\em i)}-- the luminosity and shape provide robust constraints on
	the chemical profile inside RG structures and demark the 
	maximum downward penetration of the convective envelope during the first 
	dredge-up (Bergbusch \& VandenBerg 2001; Cassisi et al.\ 2002); 
	{\em ii)}-- the luminosity peak can also be used to estimate 
	either the cluster metallicity (Desidera et al.\ 1998) or the 
	cluster distance. However, the bump luminosity is 
	affected by both theoretical (treatment of the convective transport 
	in stellar envelopes) and observational (distance, reddening) uncertainties. 
	To overcome the latter, Fusi Pecci et al.\ 1990, hereinafter FP90) 
	suggested using the $\Delta V_{HB}^{bump}$=$V(bump)-V(HB)$ parameter, 
	i.e., the difference in apparent
	visual magnitude between the RGB bump and the Horizontal Branch (HB) at the 
	luminosity level of RR-Lyrae instability strip. Using new estimates of this 
	parameter for 11~GCs, 
	FP90 found that the observed $\Delta V_{HB}^{bump}$ values were 0.4 mag 
	fainter than predicted. This gave rise to a series of 
	theoretical (Alongi et al.\ 1991; Bono \& Castellani 1992; 
	Bergbusch \& VandenBerg 2001) 
	and observational (Alves \& Sarajedini 1999; Ferraro et al.\ 1999) 
	investigations. 

	Cassisi \& Salaris (1997) gave a
	new spin to this problem, finding 
	good agreement between theory and observations once the global 
	metallicity of individual clusters was taken into account. This was 
	confirmed by Zoccali et al.\ (1999) using a large sample (28) of 
	GCs covering a wide metallicity range ($-2.1$$\le$[Fe/H]$\le$$-0.3$ dex).
	The comparison between 
	the evolutionary lifetimes during the crossing of the H-discontinuity and
	star counts across the RGB bump led also to good agreement between theory and 
	observations (Bono et al.\ 2001). 
	More recently, Riello et al.\ (2003, hereinafter R03) found that predicted 
	$\Delta V_{HB}^{bump}$ 
	values agree quite well with observations for cluster ages of 12$\pm$4~Gyr. 
	This investigation used a sample of 54 GCs, but only three metal-poor
	([M/H]$\le$--1.75 dex) GCs were included. Note that identification of the 
	RGB bump in metal-poor GCs is more difficult since it is
	brighter than the HB, where evolution along the RGB becomes faster and the
	stellar sample, consequently, becomes smaller.    
	 
	Here we augment the previous observations by investigating $\Delta V_{HB}^{bump}$ 
	in a large sample of GCs with metallicities  ranging from $-2.43$ to $-0.70$ dex.  
	In particular, we focus our attention on different sources of possible 
	systematic errors, namely metallicity scale, $\alpha$ elements, 
	CNO abundances and helium content.

	%------------------------------------------------------------------------
	\section{Empirical and theoretical frameworks}
	From the database maintained by Stetson (2000) we selected 15~GCs with
	low reddening ($E(\bmv)\,\le\,0.10$, Harris 1996) and 
	a wide range in metallicity (--2.43$\le$[Fe/H]$\le$--0.70, 
	see data listed in Table~1).
	The $B$ and $V$ data adopted in this investigation come from 
	original and archival observations which have been collected,
	reduced, and calibrated by PBS in an ongoing effort 
	to provide homogeneous photometry on the Landolt (1992) photometric 
	system for a significant fraction of GCs. More details concerning 
	individual fields can be found at the following URL: 
	http://www4.cadc-ccda.hia-iha.nrc-cnrc.gc.ca/community/STETSON/standards/ 
	or by direct communication with PBS.

	To find the RGB bump we represented each RG star detected in the 
	$B,V$-bands by a Gaussian kernel with a $\sigma$ equal to the photometric 
	uncertainty. The differential RGB LF was formed by summing the individual 
	Gaussians and the position of the bump was estimated by fitting the resulting
	peak with a Gaussian. 
	% Point C 
	Data plotted in Fig.~2 show that the photometric error in color, added
	in quadrature, is on average of the order of 0.01 mag from
	bright RGs to stars fainter than the main sequence turn-off. Therefore,
	the uncertainty in the measured position was assumed, as a generous
	estimate, equal to the sigma of the fitting Gaussian.
%AA
        The fit was performed using an interactive program that performs analytical 
	fits to the main peaks of a given distribution (Calamida et al.\ 2009). On 
	the basis of the residuals between the RGB LF and the analytical fit, the 
	software allows us to remove/insert new Gaussian components manually. 

	%%%%%%%%%%%%%%%%%%%%%%%%%%%%%%%%%%%%%%%%%%%%%%%%%%%%%%%%%%%%%%%%%%%%%%%%%%%%%%%%%%%%%
	% 			fig 1 
	%%%%%%%%%%%%%%%%%%%%%%%%%%%%%%%%%%%%%%%%%%%%%%%%%%%%%%%%%%%%%%%%%%%%%%%%%%%%%%%%%%%%%
	\begin{figure}[!ht]
	\begin{center}
	\label{fig1}
	\includegraphics[height=0.45\textheight,width=0.450\textwidth]{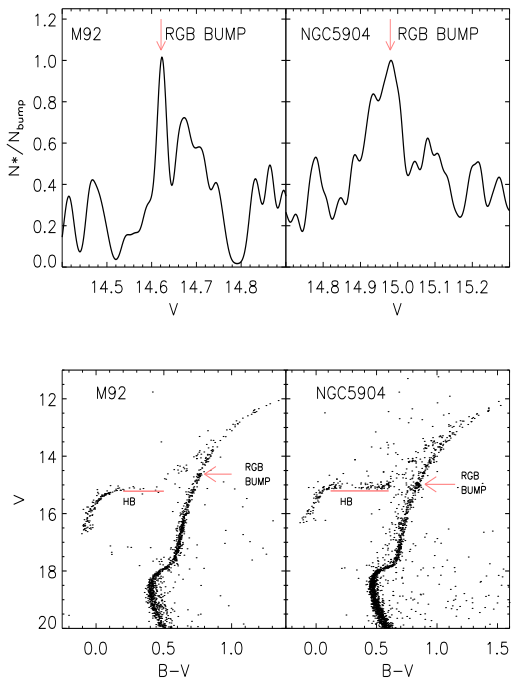}
	\vspace*{-0.15truecm}
	\caption{Top -- RGB luminosity function for a metal-poor (M92, 
	[Fe/H]=--2.38 dex, left) and a metal-intermediate (NGC~5904, 
	[Fe/H]=--1.32 dex, right) GC. The star counts ($N_\star$) are 
	normalized to the number of stars in the RGB bump. The arrows mark 
	the position of the RGB bump. 
	Bottom -- $(V,\bmv)$ CMDs of the two GCs plotted in the top panels. 
	The thin solid lines show the HB level, while the arrows indicate the RGB bump. 
	}
	\end{center}
	\end{figure}
	%%%%%%%%%%%%%%%%%%%%%%%%%%%%%%%%%%%%%%%%%%%%%%%%%%%%%%%%%%%%%%%%%%%%%%%%%%%%%%%%%%%%%
	% 			fig 2 
	%%%%%%%%%%%%%%%%%%%%%%%%%%%%%%%%%%%%%%%%%%%%%%%%%%%%%%%%%%%%%%%%%%%%%%%%%%%%%%%%%%%%%
	\begin{figure}[!ht]
	\begin{center}
	\label{fig2}
	\includegraphics[height=0.3\textheight,width=0.450\textwidth]{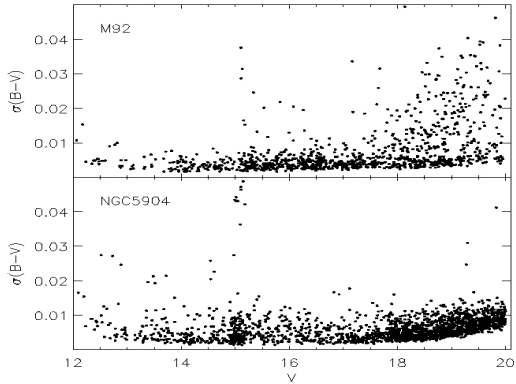}
	\vspace*{-0.15truecm}
	\caption{Intrinsic photometric error in color (\bmv) as a function of 
	$V$ magnitude for M92 (top) and NGC5904 (bottom). In each cluster the outliers
	at $V\,\sim\,15$ are RR Lyrae variable stars.    
	}
	\end{center}
	\end{figure}

	The HB level for GCs with either an intermediate or a red HB morphology was 
        estimated using the LF of HB stars in the flat region. Specifically, the 
        ZAHB level was estimated at 3$\sigma$ fainter than the peak of the LF 
        (see bottom right panel of Fig.~1). 
% Point D ter 
        Note that the 3$\sigma$ values range from $\sim$0.09 in metal-poor clusters
        to $\sim$0.18 mag in metal-rich clusters.  
	% Point D 
	This approach may be compared to the procedure adopted by Zoccali et al.\ (1999)
	and by Riello et al.\ (2003) for GCs with a red HB morphology (metal-rich).
	They fit the lower envelope of the HB stellar distribution and the ZAHB level 
        was fixed at $3\sigma$ above the lower envelope. The former approach is  
        less affected by accidental measuring errors, since it relies on the peak 
        of the LF.  
	On the other hand, for GGCs with a blue HB morphology (typically metal-poor
	and metal-intermediate GCs with only a few red HB stars) we used a template
	cluster with the same metallicity and a well populated HB. The HB level
	was estimated by shifting the template cluster in magnitude and in color
	(Buonanno et al.\ 1986; Ferraro et al.\ 1992; Zoccali et al.\ 1999).
	The observational errors for the CMDs of Fig.~1 are presented in Fig.~2. 
	The uncertainty of $\Delta V_{HB}^{bump}$ was estimated by adding in 
	quadrature the standard errors of the bump and HB luminosities (see Table~1). 
	To increase the sample size we also included the $\Delta V_{HB}^{bump}$ 
	of R03. However, their $\Delta V_{HB}^{bump}$ parameters 
	were based on the {\rm F555W}-band of WFPC2@HST (see their 
	Table~1). We have nine clusters in common with R03 and we found 
	a mean difference of $0.06\pm0.03$, i.e., a factor of two smaller 
	than the typical uncertainty. We applied this shift in magnitude to 
        the R03 estimates; our results are independent of this magnitude correction.       
	For the clusters in common with R03, we adopted our $\Delta V_{HB}^{bump}$ 
	values, since the number of RGB stars across the bump region is larger. 
	We ended up with a sample of 62 GCs.  

	To compare theory and observations we adopted different metallicity scales.
	1)~The metallicity scale by Carretta et al.\ (2009, hereinafter C09), based
	on iron measurements of Fe{\footnotesize I} lines from intermediate resolution 
	($R\approx$ 20,000) spectra collected with GIRAFFE at the Very 
	Large Telescope (VLT) and from high-resolution spectra collected with 
	UVES at VLT ($R\approx$ 40,000). They observed 19 calibrating GGCs and for 
	each cluster they collected $\sim$100 GIRAFFE and $\sim$10~UVES spectra. 
	The estimated mean uncertainty of these abundances ranges from 
	0.03 (NGC~1904) to 0.08 (NGC~2808) dex. 
	2)~The metallicity scale by Kraft \& Ivans (2003,2004, 
	hereinafter KI03 and KI04) based on high-resolution spectra 
        ($R\approx$ 30,000, Shetrone \& Keane 2000) and on MARCS atmosphere 
        models.  This scale has two key advantages: 
	{\em i)}~iron abundances based on measurements of Fe{\footnotesize II} lines, 
	which are minimally affected by non-LTE
	effects (Thevenin \& Idiart 1999); and {\em ii)}~robust determinations 
	of surface gravity and effective temperature. The accuracy of this scale 
	is on average better than 0.1 dex. 
	3)~Metallicities by Rutledge et al.\ (1997, hereinafter RHS97), derived 
	from homogeneous and accurate measurements of the calcium triplet (CaT), 
	referred to both the Zinn \& West (1984, hereinafter ZW84) and the 
	Carretta \& Gratton (1997, hereinafter CG97) metallicity scales.
	The typical precision of these metallicities is 0.10--0.15 dex (KI04).
	Note that the cluster metallicities provided by ZW84 were based on various 
	diagnostics (G band, Ca{\footnotesize II} K, and Mg{\footnotesize I} lines), 
	and their typical precision is $\sim$0.15 dex.

	%%%%%%%%%%%%%%%%%%%%%%%%%%%%%%%%%%%%%%%%%%%%%%%%%%%%%%%%%%%%%%%%%%%%%%%%%%%%%%%%%%%%%
	% 			fig 3
	%%%%%%%%%%%%%%%%%%%%%%%%%%%%%%%%%%%%%%%%%%%%%%%%%%%%%%%%%%%%%%%%%%%%%%%%%%%%%%%%%%%%%
	\begin{figure}[!ht]
	\begin{center}
	\label{fig3}
	\includegraphics[height=0.50\textheight,width=0.50\textwidth]{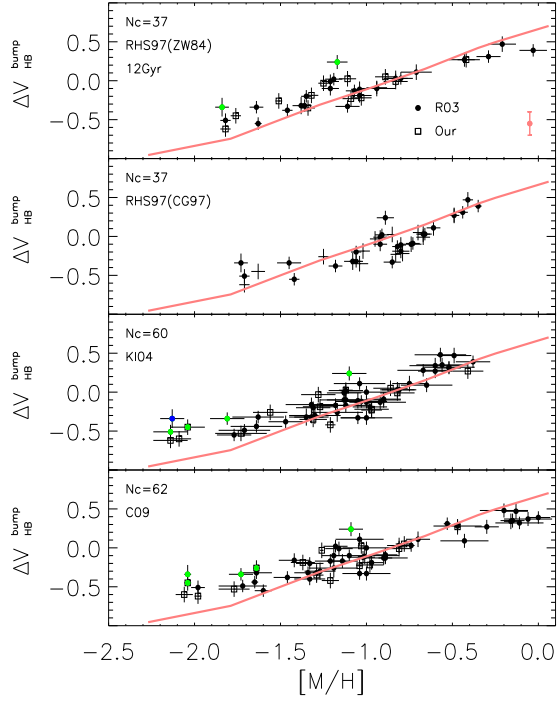}
	%\vspace*{-0.05truecm}
	\caption{$\Delta V_{HB}^{bump}$ as a function of the global metallicity 
	[M/H]. From top to bottom the same data are plotted using different 
	metallicity scales: 
	Rutledge et al. (1997, RHS97) in the Zinn  \&  West (1984, ZW84) scale; 
	RHS97 in the Carretta \& Gratton (1997, CG97) scale; 
	Kraft \& Ivans (2004, KI04); C09. 
	The number of GCs in each panel is also labeled ($N_{c}$). 
	The circles display $\Delta V_{HB}^{bump}$ parameters from 
	Riello et al.\ (2003), while the squares show current estimates.  
	The red line shows predictions (BaSTI) at fixed age (12~Gyr) for $\alpha$-enhanced chemical compositions. 
	The error bar plotted in the right corner of the top panel shows 
	theoretical uncertainties. The GCs with difference from theory 
	larger than 2$\sigma$ and 3$\sigma$ are plotted in green and 
	in blue. 
	}
	\end{center}
	\end{figure}

	To estimate the global metallicity ([M/H]) of individual GCs, 
	we adopted the relation by Salaris et al.\ (1993) and assumed
	[$\alpha$/Fe]=+0.4 dex for all GCs. 
%DD1    
        This assumption is based on spectroscopic measurements of $\alpha$-elements 
	in GCs (see Fig.~4 in Gratton et al.\ 2004). 
%DD1 ter 
        The idea of a constant [$\alpha$/Fe] as a function of the metal 
        content relies only on three metal-rich GGCs (NGC~6528, NGC~6553, Liller~1), 
        while field stars show a steady decrease in $\alpha$-element abundances in 
        the approach to solar chemical composition. However, we adopted a constant
        [$\alpha$/Fe] ratio, because field and cluster stars show different abundance
        patterns (see \S 3).   
        Among our 62 GCs, metallicity estimates are available for 37 objects 
        in RHS97 and 60 in KI04, while the entire sample is included in the 
        C09 metallicity compilation.

	To compare theory with observations we adopted the $\alpha$-enhanced 
	isochrones and Zero Age Horizontal Branches (ZAHBs) available in the 
	BaSTI database\footnote{Evolutionary models can be download from
	http://www.oa-teramo.inaf.it/BASTI}. 
	We adopted isochrones that neglect atomic diffusion 
	(Pietrinferni et al.\ 2006). The global metallicity ranges from 
	[M/H]=--2.27 to +0.06 dex, the primordial helium content by mass 
	is $Y_p$=0.245, and the helium-to-metal enrichment ratio, 
	$\Delta Y$/$\Delta Z$, is 1.4, while the stellar mass of the 
	progenitors (Main Sequence Turn Off, MSTO) ranges from 
	0.78 $M_\odot$ ([M/H]=--2.27) to 0.90 $M_\odot$ ([M/H]=+0.06).  
	Predicted $\Delta V_{HB}^{bump}$ values were estimated assuming a GC age 
	of 12~Gyr (VandenBerg et al.\ 2006). 

	To estimate the typical uncertainty in the predicted  $\Delta V_{HB}^{bump}$
	values we assumed an uncertainty of 0.1 dex in the heavy element abundances 
	(Fe, $\alpha$-elements), 0.02 in the helium content by mass 
	($Y$), and 1~Gyr in the cluster age. Moreover, we also included 
	uncertainties in the input physics used in constructing evolutionary 
	models (Cassisi et al.\ 1998; Bergbusch \& VandenBerg 2001).

	%%%%%%%%%%%%%%%%%%%%%%%%%%%%%%%%%%%%%%%%%%%%%%%%%%%%%%%%%%%%%%%%%%%%%%%%%%%%%
	\section{Comparison with $\alpha$- and CNO-enhanced models}
	Fig.~3 shows the comparison between empirical and predicted 
	$\Delta V_{HB}^{bump}$ values in the four different metallicity scales.   
	The figure shows that the observed 
	$\Delta V_{HB}^{bump}$ values of metal-poor and metal-intermediate 
	GCs are systematically larger than predicted. This discrepancy does 
	not depend on the adopted metallicity scale and becomes strongest 
	in metal-poor ([M/H]$\le$--1.5) GCs.  
	To be more quantitative we calculated for each GC the quadrature sum
	of theoretical and empirical uncertainties (see Table~1 and 
	Fig.~3). 
	We found that the GCs with a discrepancy larger than
	$2\sigma$ (green symbols in Fig.~3) are two, four and five for the 
	RHS97[ZW84], the KI04 and the C09  metallicity scale. It is worth 
	mentioning that the difference between theory and observations is 
	$\approx$0.40 mag.  
	We also found that one GC in KI04 shows a discrepancy larger 
	than $3\sigma$ ($\approx$0.60 mag). This suggests that 
	metallicities based either on KI04 or on C09 scale show 
	a very similar behavior when compared with observations.  
	However, the bottom panel shows that several metal-rich 
	([M/H]$\approx$--0.5 dex) GCs in the C09 metallicity scale 
	are located below the predicted loci. Note that possible 
	evolutionary effects in the HB phase would imply an increase 
	(more positive) in the value of the $\Delta V_{HB}^{bump}$ parameter.   
	Fig.~3 indicates that---if we trust the KI04 and C09 metallicity 
	scales---approximately 40\% of the metal-poor ([M/H]$\le$--1.7, --1.6 dex) 
	GCs show a discrepancy with theory that is at least at a
	$2\sigma$ level.  Such a discrepancy can hardly be explained 
	as a spread in cluster age:  the age derivative of the 
	$\Delta V_{HB}^{bump}$ parameter is $\approx$ 0.03 mag per Gyr over 
	the entire metallicity range (see, e.g., Cassisi \& Salaris 1997; 
	Zoccali et al.\ 1999), and an age spread of order 10~Gyr seems out
	of the question.  Moreover, changes in mixing-length 
	efficiency have little effect on the luminosity of the RGB bump: 
	a change of $\pm$0.1 in ($\alpha_{ml}$) 
	causes a variation of $\sim$0.03 mag in the $\Delta$V$^{bump}_{HB}$ 
	parameter (Cassisi \& Salaris 1997). The comparison between predicted 
	and empirical cluster RGBs does not support larger variations in 
	the $\alpha_{ml}$ parameter.  
%DD
        The anonymous referee raised the problem that we are using a sample of 
	$\Delta V_{HB}^{bump}$ parameters that are based on two different methods 
	to estimate the ZAHB luminosity. According to the referee {\em there is a big 
	possibility of systematic errors in the ZAHB determinations}. However, 
	some circumstantial evidence argues against this working hypothesis.\\
	{\em i)}-- The photometric precision of the current data and of the HST data adopted 
	by Riello et al.\ at the ZAHB level is better than 0.010-0.015 mag over the 
	entire cluster sample. We have nine GCs in common with Riello et al. that 
	cover the entire metallicity range and the difference in the estimated 
	$\Delta V_{HB}^{bump}$ parameters is smaller than 0.1 mag.\\ 
	{\em ii)}-- The method we adopted to estimate the ZAHB luminosity (3$\sigma$ 
	fainter than the peak of HB LF) might be prone to underestimate the real ZAHB 
	level. However, a less conservative estimate would imply brighter ZAHBs, and 
	in turn larger (more positive) $\Delta V_{HB}^{bump}$ parameter and therefore 
	a larger discrepancy between theory and observations.\\   
	{\em iii)}-- We estimated the $\Delta V_{HB}^{bump}$ parameter for a good 
	fraction (six out of eleven) of metal-poor ([M/H]$\lesssim$--1.7, --1.6 dex) 
	GCs. If we do not include the GCs measured by Riello et al. the 
	discrepancy between theory and observations, in this metallicity range, 
	becomes larger.\\ 
        {\em iv)}-- The majority (nine out of twelve) of the $\Delta V_{HB}^{bump}$ 
	parameters of metal-rich ([M/H]$\ge$--0.3 dex) GCs have been estimated by 
	Riello et al.\\     
	{\em v)}-- To overcome subtle uncertainties in empirical estimates and 
	theoretical predictions we are focusing our attention on discrepancies 
	at the level of several tenths of a magnitude.\\
	
	%%%%%%%%%%%%%%%%%%%%%%%%%%%%%%%%%%%%%%%%%%%%%%%%%%%%%%%%%%%%%%%%%%%%%%%%%%%%%%%%%%%%%
	% 			fig 4
	%%%%%%%%%%%%%%%%%%%%%%%%%%%%%%%%%%%%%%%%%%%%%%%%%%%%%%%%%%%%%%%%%%%%%%%%%%%%%%%%%%%%%
	\begin{figure}[!ht]
	\begin{center}
	\label{fig4}
	\includegraphics[height=0.350\textheight,width=0.450\textwidth]{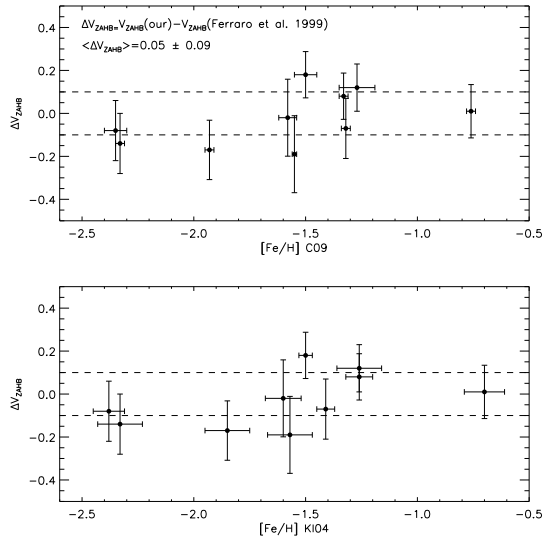}
	%\vspace*{-0.05truecm}
	\caption{
        Top -- Difference between current $V_{ZAHB}$ estimates and similar estimates 
        by Ferraro et al. (1999) as a function of the iron abundance (C09). The 
        dashed lines display the uncertainty ($\pm0.1$ mag) we adopted for 
        $V_{ZAHB}$ estimates. The error bars account for individual uncertainties on 
        $V_{ZAHB}$ (sum in quadrature) and in iron abundance. Bottom -- Same as the top, 
        but for the iron abundaces based on the metallicity scale by KI04.  
	}
	\end{center}
	\end{figure}

% DD ter 
        Finally, we also compared current estimates of $V_{ZAHB}$ with similar 
        estimates provided by Ferraro et al.\ (1999) to test whether our approach 
        {\em may also introduce a zero-point shift or a systematic trend with 
        metallicity}. The quoted authors to estimate $V_{ZAHB}$, in a large sample 
        of GGCs with non homogeneous photometry, applied a metallicity dependent 
        correction to $V_{HB}$ ranging from 0.06 to 0.16 mag. We have ten GGCs 
        in common with Ferraro et al. and data plotted in Fig.~4 show no evidence 
        of a trend with metallicity and the two different sets of $V_{ZAHB}$ 
        estimates agree within 1$\sigma$. 
	 
	There are several culprits that can explain the difference between 
	current predictions and those provided by Cassisi \& Salaris (1997).
	Their primordial He-content was $Y$=0.23 while ours is $Y$=0.245, and 
	the estimated cluster age has been decreased from 15 to 12~Gyr. The input 
        physics (opacities, equation of state) was also changed, and they adopted 
	a scaled-Solar heavy-element mix rather than an $alpha$-enhanced one. 
        However, the difference between 
	the old and the new scaled-solar $\Delta V_{HB}^{bump}$ parameter for 
	[M/H]=--1.79 and t=15~Gyr is small (--0.65 vs --0.60). The 
	difference with the new $\alpha$-enhanced $\Delta V_{HB}^{bump}$ parameter, 
	for the same chemical composition and an age of 12~Gyr, is also modest 
	($-0.65$ vs $-0.75$). 
	% Point E 
	Given the limited changes between the current and previous
	predictions, the discrepancy between theory and observations
	in the metal-poor tail is confirmed and reinforced by the inclusion of
	new GCs in this metallicity regime.

	%%%%%%%%%%%%%%%%%%%%%%%%%%%%%%%%%%%%%%%%%%%%%%%%%%%%%%%%%%%%%%%%%%%%%%%%%%%%%%%%%%%%%
	% 			fig 5
	%%%%%%%%%%%%%%%%%%%%%%%%%%%%%%%%%%%%%%%%%%%%%%%%%%%%%%%%%%%%%%%%%%%%%%%%%%%%%%%%%%%%%
	\begin{figure}[!ht]
	\begin{center}
	\label{fig4}
	\includegraphics[height=0.450\textheight,width=0.450\textwidth]{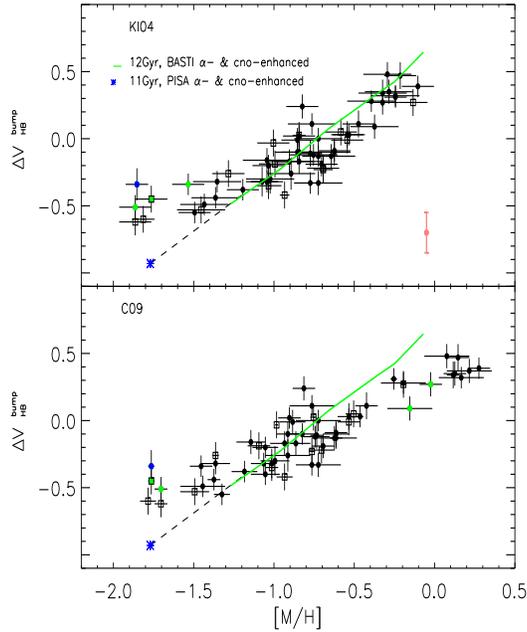}
	%\vspace*{-0.05truecm}
	\caption{Same as Fig.~3, but for two different metallicity scales: 
	Kraft \& Ivans (2003, KI04, top) and Carretta et al.\ (2009, C09, bottom).
	The green line shows  predictions (BaSTI) at fixed age (12~Gyr) and
	for $\alpha$- and CNO-enhanced chemical compositions 
	(--0.07$\le$[M/H]$\le$--1.27 dex). The blue asterisk marks the predicted 
	$\Delta V_{HB}^{bump}$ parameter for an age of 11~Gyr and a 
	metal-poor ([M/H]=--1.77 dex) $\alpha$- and CNO-enhanced 
	chemical composition. The dashed line shows a linear interpolation 
	between the BaSTI predictions and the new computed metal-poor value.     
	}
	\end{center}
	\end{figure}
	 
	% Point G 
	We note in passing that theoretical models indicate that the RGB bump
	luminosity is marginally affected by plausible assumptions concerning
	the smoothing of the chemical discontinuity (Bono et al.\ 1992). Following
	a different approach, Alongi et al.\ (1991) found that the inclusion of
	nonlocal overshoot at the base of the convective envelope causes,
%DD2                           
	in low-mass RGB stars, an increase of 0.4 $V$~mag in the location of the
	RGB bump. However, this systematic shift minimally depends on the
	chemical composition and applies to both metal-poor and metal-rich
	stellar structures.

	Spectroscopic measurements suggest variations in the 
	abundance pattern in GCs on a star-by-star basis. Together with 
	changes in the relative abundances of CNO elements, 
	well defined anti-correlations have been found between 
	{\rm O} and {\rm Na} and between {\rm Mg} and {\rm Al} 
	(Gratton, Sneden, \& Carretta 2004, and references therein).       
	This applies not only to evolved and unevolved
	cluster stars (Briley et al. 1996; Cohen \& Melendez 2005) 
	but also to RG stars characterized by different thicknesses
	of the convective envelope (Pietrinferni et al.\ 2009). 
	This solid result and the recent evidence that field halo stars 
	do not show, at fixed iron content, similar changes in (O,Na,Mg,Al) 
	suggests the occurrence of an intra-cluster pollution mechanism
	(Denissenkov \& Weiss 2004; Ventura \& D'Antona 2005; 
	Salaris et al.\ 2006). Moreover, Yong et al.\ (2009) found that 
	the [C+N+O/Fe] abundance ratio changes by a factor of four 
	(0.6 dex) among bright RGs in the GC NGC~1851. This is consistent with
	the scenario suggested by Cassisi et al.\ (2008) to explain 
	the two subgiant branches detected in this cluster (Milone et al.\ 2008) 
	as a difference in the mixture of {\rm C+N+O} elements.       

	%%%%%%%%%%%%%%%%%%%%%%%%%%%%%%%%%%%%%%%%%%%%%%%%%%%%%%%%%%%%%%%%%%%%%%%%%%%%%%%%%%%%%
	% 			fig 5
	%%%%%%%%%%%%%%%%%%%%%%%%%%%%%%%%%%%%%%%%%%%%%%%%%%%%%%%%%%%%%%%%%%%%%%%%%%%%%%%%%%%%%
	\begin{figure}[!ht]
	\begin{center}
	\label{fig5}
	\includegraphics[height=0.450\textheight,width=0.450\textwidth]{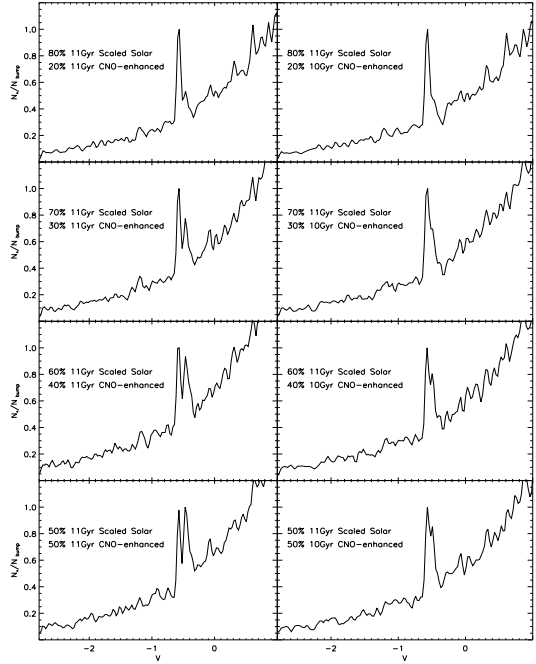}
	%\vspace*{-0.05truecm}
	\caption{Synthetic RGB luminosity functions for a metal-poor ([Fe/H]=--2.32, 
        [$\alpha$/Fe]=0.3, $Y$=0.246) GC. The total number 
	of synthetic stars is 2,500 per numerical simulation. From top to bottom the cluster 
	stellar content includes fractions of 20\%, 30\%, 40\% and 50\% of CNO-enhanced stars.
	Simulations were performed assuming either the same age (11~Gyr) for the canonical and 
	the CNO-enhanced sub-populations (left panels) or a difference of 1~Gyr between the 
	two sub-populations (right panels). 
	}
	\end{center}
	\end{figure}

	Therefore, to investigate the discrepancy between theory 
	and observations of $\Delta V_{HB}^{bump}$, 
	we also considered cluster isochrones and ZAHBs constructed 
	assuming CNO-Na abundance anti-correlations (Pietrinferni et al.\ 2009). 
	The mixture adopted in these evolutionary models includes a 
	sum in CNO abundances that is a factor of two larger than 
	in a canonical $\alpha$-enhanced mixture and within the observed 
	range of GC anti-correlations (Salaris et al.\ 2006).  
	Predictions by Pietrinferni et al.\ (2009) are  
	available for stars with metallicities
	from [M/H]=--1.27 ($M_{MSTO}$=0.79$M_\odot$) to 
	[M/H]=--0.07 ($M_{MSTO}$=0.83$M_\odot$) dex. 
	To constrain the predictions in the metal-poor regime, 
	we used an updated version of the FRANEC evolutionary 
	code (Cariulo et al.\ 2004; Degl'Innocenti et al.\ 2008) 
	to compute stellar isochrones and ZAHBs for 
	[M/H]=--1.77 ([Fe/H]=--2.32, [$\alpha$/Fe]=0.3,$Y$=0.246, $M_{MSTO}$=0.78$M_\odot$) 
	accounting for both $\alpha$- and CNO-enhancement. 
	These models include atomic diffusion, which causes a 
	decrease in the inferred cluster age of $\sim$1~Gyr (Bahcall \& 
	Pinsonneault 1995; Castellani et al.\ 1997). Therefore, 
	we estimated the $\Delta V_{HB}^{bump}$ parameter for a cluster 
	age of 11~Gyr and the blue asterisk marks its position in Fig.~5.
%DD3 
        A more detailed discussion concerning the input physics of these models 
	will be addressed in a forthcoming investigation (Di Cecco et al.\ 2010).
%DD4 
        Note that current predictions for CNO-enhanced models do not cover the 
	metallicity range  -1.77$<$[M/H]$<$-1.27  (see dashed line in Fig~5). However, 
	the linear interpolation (see dashed line in Fig.~5) is supported by the      
	smooth change showed by more metal-rich structures (Pietrinferni et al.\ 2009). 
	To compare theory and observations, the global metallicity of 
	GCs was estimated using the Salaris et al.\ (1993) relation 
	with [$\alpha$/Fe]=0.4 and [CNO/Fe]=0.3.   
	The comparison indicates that $\alpha$- and CNO-enhanced 
	predictions display a similar  discrepancy with empirical estimates 
        in the KI04 (three GCs with a difference larger than 2$\sigma$, and 
        one with a difference larger than 3$\sigma$) and C09 (four GCs with a 
        difference larger than 2$\sigma$, and one with a difference larger 
        than 3$\sigma$) metallicity scale (see the green lines in Fig.~5). 
        The difference is mainly due to the fact that the RGB bump is more 
        sensitive to an increase in CNO abundances than the ZAHB (see Fig.~1 
        and 3 in Salaris et al.\ 2008). On the other hand, the bottom 
	panel of Fig.~5 shows that, if we adopt the C09 metallicity scale, 
	a group of six GCs in the metal-rich regime ([M/H]$\ge$0.0)
	shows systematically smaller (more negative) $\Delta V_{HB}^{bump}$ 
	values than predicted. Current predictions do not cover this metallicity 
	range and we linearly extrapolated the trend from more metal-poor structures.   
	Note that Fig.~5 shows only the KI04 and the C09 metallicity scales 
	because they include the largest fraction of GCs in our sample.  
	The main outcome of the above comparison is that the observed 
	$\Delta V_{HB}^{bump}$ parameters in the metal-poor regime are 
	systematically larger (more positive), independent of the 
	metallicity scale, than predicted by $\alpha$- and CNO-enhanced 
	models. A similar discrepancy, but with observed $\Delta V_{HB}^{bump}$ 
	parameters systematically smaller than predicted,  might be also present 
	in the metal-rich tail ([M/H]$\ge$0.0), if we adopt the C09 metallicity 
	scale.          

	% Point F 
	The anonymous referee suggested a test to constrain the
	fraction of CNO-enhanced stars that could be identified as a secondary
	RGB bump. Accordingly, we computed a series of synthetic
	CMDs assuming a metal-poor chemical composition ($Y$=0.246, [Fe/H]=-2.32,
	[$\alpha$/Fe]=0.3, Pisa evolutionary code) and a cluster age
	of 11~Gyr. Together with this canonical stellar population we also
	included different fractions of a CNO-enhaced stellar component.
	We assumed for the CNO-enhanced population two different ages,
	namely 11 and 10~Gyr (see the left and the right panels of Fig.~6).
	Moreover, we have made our best efforts to match the synthetic CMD 
        to the observed CMD of M92. In particular, we randomly distributed the 
        same number of stars (2,500, $V\ge$ 16 mag) along the predicted RGB to 
        mimic the Poisson uncertainties properly and accounting for photometric 
        errors ($\sigma_B,\sigma_V=\le$ 0.015 mag).
	Fig.~6 shows the predicted LFs with CNO-enhanced fractions
	from 20\% (top) to 50\% (bottom).  To define the
	RGB bump we adopted the same procedure as for observed GCs.
	Data plotted on the left column (coeval populations) indicate that
	CNO-enhanced stars show up as a secondary
	RGB bump when its fraction is equal to or larger than 30\%. Note that
	the two peaks differ in magnitude by 0.1 mag, i.e. an order of magnitude
	larger than the typical photometric error. On the other hand, if we
	assume an age difference of 1~Gyr between the canonical and the
	CNO-enhanced population (right column) the secondary population
	shows up as a second RGB bump only when its fraction becomes
	of the order of 40-50\%. This is due to the fact that the effects
	of a decrease in cluster age and of an increase in the global metallicity
	almost cancel out across the RGB bump. The magnitude difference
	between the peaks is 0.08 mag. This indicates that RGB bump is not a good
	indicator of the presence of a sub-population that is both CNO-enhanced and
	younger relative to the main population.

	%%%%%%%%%%%%%%%%%%%%%%%%%%%%%%%%%%%%%%%%%%%%%%%%%%%%%%%%%%%%%%%%%%%%%%%%%%%%%
	\section{Comparison with He-enhanced models}

	%%%%%%%%%%%%%%%%%%%%%%%%%%%%%%%%%%%%%%%%%%%%%%%%%%%%%%%%%%%%%%%%%%%%%%%%%%%%%%%%%%%%%
	% 			fig 6
	%%%%%%%%%%%%%%%%%%%%%%%%%%%%%%%%%%%%%%%%%%%%%%%%%%%%%%%%%%%%%%%%%%%%%%%%%%%%%%%%%%%%%
	\begin{figure}[!ht]
	\begin{center}
	\label{fig6}
	\vspace*{-0.7 truecm}
	\includegraphics[height=0.500\textheight,width=0.50\textwidth]{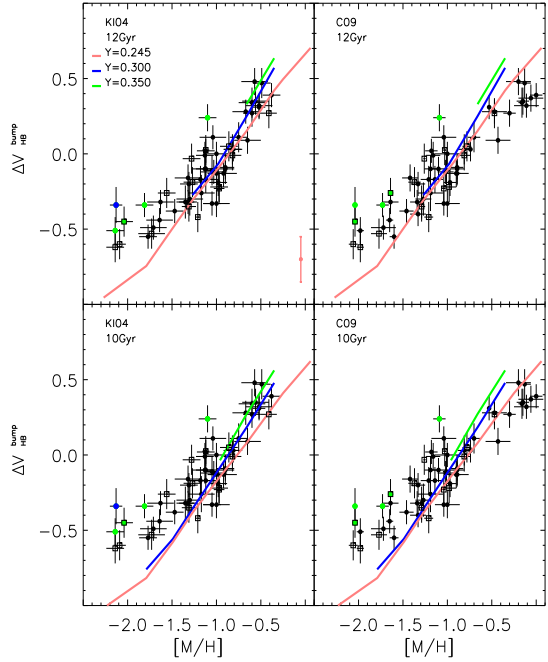}
	\caption{Same as Fig.~3, but for two different metallicity scales: 
	Kraft \& Ivans (2004, KI04, left) and Carretta et al.\ (2009, C09, right).
	Top -- The lines display predictions (BaSTI) at fixed cluster age (12~Gyr), 
	$\alpha$-enhanced chemical composition and different helium 
	contents: $Y$=0.245 (red), $Y$=0.30 (blue) and $Y$=0.35 (green). 
	Bottom -- Same as the top, but the predictions refer to cluster 
	isochrones of 10~Gyr. 
	}
	\end{center}
	\end{figure}

	Recently, accurate and deep photometry has revealed
	multiple stellar populations in several massive GCs.
	Together with the highly complex system $\omega$ Centauri (Anderson 2002; 
	Bedin et al.\ 2004) multiple stellar sequences have been detected in 
	GCs covering a broad range of metal content:
	NGC~2808 (D'Antona et al.\ 2005; Piotto et al.\ 2007), 
	M54 (Siegel et al.\ 2007), NGC~1851 (Milone et al.\ 2008) 
	and M4 (Marino et al.\ 2008). Some of these multiple sequences 
	($\omega$ Cen, NGC~2808) might be explained with a He-enhanced 
	stellar population (Norris 2004; D'Antona et al.\ 2005,2008; 
	Piotto et al.\ 2007). 
	 
	To evaluate the impact that the He-content has on 
	$\Delta V_{HB}^{bump}$ we also considered 
	evolutionary models assuming higher helium 
	abundances. The top panels of Fig.~7 show the comparison, 
	at fixed cluster age (12~Gyr), between observed and predicted 
	(BaSTI) $\Delta V_{HB}^{bump}$ parameters. Together with the 
	canonical $\alpha$-enhanced models 
	(red line) the He-enhanced models are plotted with a blue 
	($Y$=0.30; $M_{MSTO}$ = 0.72$M_\odot$ at [M/H]=--1.27 and 
        $M_{MSTO}$=0.78$M_\odot$ at [M/H]=--0.35)) and a green 
	($Y$=0.35; $M_{MSTO}$ = 0.68$M_\odot$ at [M/H]=--0.66 and 
	$M_{MSTO}$=0.72 $M_\odot$ at [M/H]=--0.35) line, respectively.

	The He-enhanced models predict larger  
	$\Delta V_{HB}^{bump}$  parameters when compared with canonical 
	models. The difference is most evident in the metal-rich tail;
	it is negligible in the metal-intermediate regime because
	an increase in helium content---at fixed cluster age---gives 
	brighter ZAHBs {\it and\/} brighter RGB bumps. Note that the He-enhanced 
	models do not exist in the metal-poor regime, since these models 
	experience either no first dredge-up, or only a mild one. The 
	RGB bump vanishes together with the chemical discontinuity in the envelope.

	Finally, we also tested the impact of simultaneous changes in both 
	cluster age and helium content.  The bottom panels of Fig.~7 show 
	the comparison between canonical and He-enhanced models for a 
	cluster age of 10~Gyr ({\bf $Y$=0.245}, 
	$M_{MSTO}$=0.82$M_\odot$ for [M/H]=--2.27 and $M_{MSTO}$=0.93$M_\odot$ 
	for [M/H]=+0.06; {\bf $Y$=0.30}, 
	$M_{MSTO}$=0.75$M_\odot$ for [M/H]=--1.79) and $M_{MSTO}$=0.81$M_\odot$ 
	for [M/H]=--0.35; {\bf $Y$=0.35}, 
	$M_{MSTO}$=0.69$M_\odot$ for [M/H]=--0.96 and  $M_{MSTO}$=0.74$M_\odot$ 
	for [M/H]=--0.35). 
	Once again the predicted $\Delta V_{HB}^{bump}$ values are systematically
	smaller than observed, regardless of the metallicity scale. The decrease in 
	cluster age expands the range in metal content within which the 
	He-enhanced models display the RGB bump. However, in the metal-intermediate 
	regime (--1.7$\lesssim$[M/H]$\lesssim$-0.8) the difference between 
	canonical and He-enhanced ($Y$=0.30) models is negligible. This 
	is because the He-enhancement causes similar changes in 
	$V_{HB}$ and $V_{bump}$ that cancel out in the difference.

	%%%%%%%%%%%%%%%%%%%%%%%%%%%%%%%%%%%%%%%%%%%%%%%%%%%%%%%%%%%%%%%%%%%%%%%%%%%%%%%%%%%%%
	% 			fig 7
	%%%%%%%%%%%%%%%%%%%%%%%%%%%%%%%%%%%%%%%%%%%%%%%%%%%%%%%%%%%%%%%%%%%%%%%%%%%%%%%%%%%%%
	\begin{figure}[!ht]
	\begin{center}
	\label{fig6}
	\vspace*{-0.7 truecm}
	\includegraphics[height=0.500\textheight,width=0.50\textwidth]{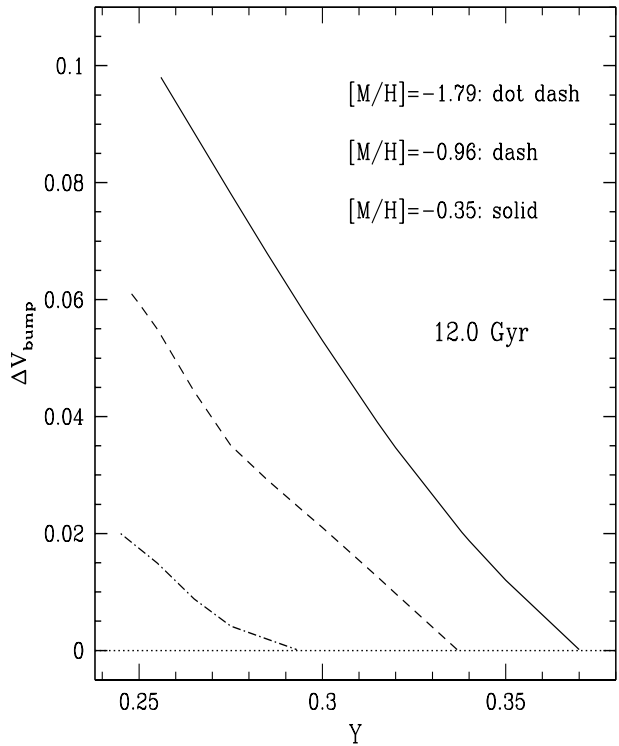}
	\caption{Predicted width in visual magnitude of the RGB bump for 
	metal-poor (dashed-dotted line), metal-intermediate (dashed line) and 
	metal-rich (solid line) stellar structures computed at fixed cluster 
	age and global metallicity, but assuming different helium contents  
	(Teramo evolutionary code).  
	}
	\end{center}
	\end{figure}

	% Point H 
	The anonymous referee suggested a test to constrain the impact
	of a sub-population of He-enhanced stars on the shape and width of the
	RGB bump. Accordingly, we computed several series of evolutionary
	models (using the Teramo evolutionary code) at fixed global metallicity,
	but steadily increasing the initial He abundance from the canonical value
	($Y\sim0.25$) to the disappearance of the bump (see Table 2). 
%HH1 
        Fig.~8 shows the predicted width in visual magnitude between the 
        faintest and the brightest point attained by the RGB bump for 
        metal-poor (dashed-dotted line), metal-intermediate (dashed line) and 
        metal-rich (solid line) stellar structures constructed assuming different 
        helium contents and fixed cluster age. This figure shows
	that the width in magnitude increases by $\approx$ a factor of five
	when moving from metal-poor to metal-rich structures. In metal-poor structures
	an increase in helium of $\sim$ 10\% causes the disappearance of the
	bump, since its width becomes smaller than one hundreth of a magnitude.
	The same result occurs in metal-intermediate and metal-rich structures,
	but the required increase in helium content is of the order of 30-35\%.
%HH1 
        The disappearance of the RGB bump is a consequence of the decrease in the 
	greatest depth of the convective envelope, and in turn in the reduced amount of 
	nuclear processed material dredged-up into the outermost regions (Salaris 
	\& Cassisi 2008). 
%HH1 ter
        This also means that an increase in helium content has a stronger 
        impact on the width of the RGB bump of metal-poor structures than of 
        the metal-rich ones. 
	In order to be more quantitative on this key dependence we also computed,
	for the same structures, the increase in surface helium abundance after
	the first dredge-up (see column 6 in Table 2). 
%HH3    
        We found that in a metal-poor structure with $Y$=0.35 the amount of extra helium,
	i.e., the amount of helium dredged-up into the surface layers by the sinking of the 
	convective envelope, decreases by a factor of two when compared with structures 
	constructed assuming a canonical helium content. 
	The same outcome applies for metal-intermediate and metal-rich structures,
	but only for $Y$=0.40.

	%%%%%%%%%%%%%%%%%%%%%%%%%%%%%%%%%%%%%%%%%%%%%%%%%%%%%%%%%%%%%%%%%%%%%%%%%%%%%
	\section{Discussion}

	New electron-conduction opacities provided by Cassisi et al.\ (2007)
	have a negligible effect on the luminosity of the RGB bump, but affect 
	the size of the helium core, and in turn the HB luminosity. 
	The helium core mass of the new models decreases, independent of the 
	metallicity, by 0.006 $M_\odot$. The difference in the $V$ band  
	is $\sim$ 0.04--0.07 mag. However, this goes in the opposite 
	direction, since the $\Delta V_{HB}^{bump}$ parameter would attain 
	smaller (more negative) values.  
	    
	Interestingly, the drift 
	between theory and observations becomes even more evident 
	when moving from the metal-intermediate to the metal-poor regime.  
	To further constrain this effect, we also investigated the 
	impact of possible uncertainties in the radiative opacities 
	(Guzik et al.\ 2009). 
	In particular, we constructed two new sets of isochrones 
	and of ZAHB models for [M/H]=--2.27 ($Z$=0.0001) and 
        [M/H]=--1.79 ($Z$=0.0003) by artificially increasing 
	the radiative opacity by 5\%. 
	The $\Delta V_{HB}^{bump}$ parameters based 
	on opacity-enhanced models are smaller, but the difference 
	with canonical predictions is of the order of a few hundredths 
	of a magnitude. There are similar changes 
	in $V_{HB}$ and $V_{bump}$ magnitude that cancel 
	in the difference.          

	The current discrepancy between theory and observations is little 
	affected by the new solar heavy-element mixture provided by 
	Asplund et al.\ (2005, hereinafter A05). The new mixture has a 
	twofold impact on GC metallicities, since it affects both the iron 
	content ([Fe/H]) and the relation between the iron and the 
	global metallicity $Z$, i.e., the heavy-element abundance 
	adopted in evolutionary computations. To quantify the difference 
	between the old and new solar mixtures on the iron measurements  
	we consider the GC M92. 
	According to KI04 this GC has an iron content of [Fe/H]=$-2.38$ dex 
	assuming a solar iron abundance in number of $\log$ $\epsilon_{Fe}$=7.51 
	(Grevesse \& Noels 1993, hereinafter GN93). On the other hand, 
	the iron abundance of M92 becomes [Fe/H]=$-2.32$ dex, if we 
	adopt the new solar iron abundance $\log$ $\epsilon_{Fe}$=7.45 (A05), 
	since  [Fe/H]$_{A05}$=[Fe/H]$_{GN93}$ + 
	$\Delta \log \epsilon$ (GN93 - A05)= [Fe/H]$_{GN93}$ + 0.06 dex. 
	The new Solar mixture by A05 gives global metallicities,  
	$Z$, that at fixed iron content are approximately 30\% 
	lower than those based on the GN93 Solar 
	mixture, since $(Z/X)^{A05}_{\odot}/(Z/X)^{GN93}_{\odot}\approx$ 0.7. 
	Given the above differences, the global metallicity
	of M92, assuming an $\alpha$-enhancement of +0.4, changes from 
	$Z$=1.51 $\cdot 10^{-4}$ (GN93) to $Z$=1.19$\cdot 10^{-4}$ (A05).

	%%%%%%%%%%%%%%%%%%%%%%%%%%%%%%%%%%%%%%%%%%%%%%%%%%%%%%%%%%%%%%%%%%%%%%%%%%%%%
	\section{Summary and final remarks}

%HH4 
	We estimated the $\Delta V_{HB}^{bump}$ parameter for 15 GGCs using accurate
	and homogeneous optical databases. Combining the new empirical evaluations 
	with similar estimates provided by Riello et al.\ (2003), we ended up with
	a sample of 62 GGCs covering a very broad range in metal content
	(--2.16$\le$[M/H]$\le$--0.58 dex) and structural parameters.
	The comparison between theory and observations produced the following findings:

	{\em a)}-- The observed $\Delta V_{HB}^{bump}$ parameters, using the
	homogeneous metallicity scales for GGCs provided by Kraft \& Ivans (2004)
	and by Carretta et al.\ (2009), are larger than the predicted ones.
	In the metal-poor regime ([M/H]$\lesssim$--1.7, --1.6 dex) 40\% of GCs
	show discrepancies of $2\sigma$ ($\approx$0.40 mag) or more. Evolutionary
	models that include either $\alpha$- and CNO-enhancement or helium
	enhancement do not alleviate the discrepancy between theory and observations.

        {\em b)}-- The above discrepancy between theory and observations is not
	affected by the new Solar heavy-element mixture.

	{\em c)}-- The comparison between $\alpha$- and CNO-enhanced evolutionary
	models and observations in the Carretta et al.\ metallicity scale also
	indicates that observed $\Delta V_{HB}^{bump}$ parameters, in the metal-rich
	regime ([M/H]$\ge$0), might be systematically smaller than predicted.

	{\em d)}-- Evolutionary models predict, as expected, a strong dependence
	of the RGB bump on the helium content. Current predictions indicate that
	an increase in helium of $\sim$ 10\% causes the disappearance of the
	RGB bump in metal-poor structures. The outcome is the same in
	metal-intermediate and in metal-rich structures, but the required 
        increase in helium content is of the order of 30-35\%.

	In this investigation we assumed a constant CNO-enhancement over the entire
	metallicity range. However, spectroscopic evidence indicates that the
	CN-strength is correlated with Na, Al and Mg in metal-poor, with Na and Al,
	in metal-intermediate and with only Na in metal-rich GCs (Smith \& Wirth 1981;
	Gratton et al.\ 2004). This implies that the CNO-enhancement might depend on
	the metal abundance (Cottrell \& Da Costa 1981).    

	More detailed investigations concerning the input physics
	adopted in the computation of low-mass evolutionary models is 
	required before we can reach a firm conclusion. This also
	applies to the metallicity scale, an in particular to the occurrence 
	of possible systematic errors when moving from very metal-poor to 
	metal-rich GCs.

	\acknowledgments
        We also thank an anonymous referee for his/her comments that helped
        us to improve the readability of the manuscript. This project was
        supported by Monte dei Paschi di Siena (P.I.: S. Degl'Innocenti),
        PRIN-MIUR2007 (P.I.: G. Piotto). MZ acknowledges support by
        Fondecyt Regular \#1085278, the FONDAP Center for Astrophysics
        \#15010003 and the Basal CATA PFB-06.

%\newpage
%--------------------------------------------------------------------------------------

%%%%%%%%%%%%%%%%%%%%%%%%%%%%%%%%%%%%%%%%%%%%%%%%%%%%%%%%%%%%%%%%%%%%%%%%%%%%%%%%%%%%%
% 			tab
%%%%%%%%%%%%%%%%%%%%%%%%%%%%%%%%%%%%%%%%%%%%%%%%%%%%%%%%%%%%%%%%%%%%%%%%%%%%%%%%%%%%%
\begin{deluxetable}{llcccccc}
\rotate
\tabletypesize{\scriptsize}
\tablewidth{0pt}                       
\tablecaption{Iron abundances and $\Delta V_{HB}^{bump}$ for the 15~GCs in our sample.}
\tablehead{
\colhead{ID}&
\colhead{Alias}&
\colhead{$V^{bump}$}&
\colhead{$\Delta V_{HB}^{bump}$}&
\multicolumn{4}{c}{[Fe/H]}\\  
\colhead{}&
\colhead{}&
\colhead{}&
\colhead{}&
\colhead{RHS97(ZW84)\tablenotemark{a}}&
\colhead{RHS97(CG97)\tablenotemark{a}}&
\colhead{C09\tablenotemark{b}}&
\colhead{KI04\tablenotemark{c}}
}
\startdata
NGC~104 & 47Tuc & $14.50\pm0.04$ & $+0.27   \pm   0.11 $ & $  -0.71  \pm  0.05  $ & $  -0.78 \pm 0.02  $ & $  -0.76 \pm 0.02  $ & $ -0.70 \pm0.09$\\
NGC~288 &       & $15.45\pm0.02$ & $+0.02   \pm   0.10 $ & $  -1.40  \pm  0.05  $ & $  -1.14 \pm 0.03  $ & $  -1.32 \pm 0.02  $ & $ -1.41 \pm0.04$\\
NGC~1261&       & $16.62\pm0.05$ & $-0.22   \pm   0.11 $ & $  -1.32  \pm  0.06  $ & $  -1.08 \pm 0.04  $ & $  -1.27 \pm 0.08  $ & $ -1.26 \pm0.10$\\ 
NGC~4590& M68   & $16.02\pm0.04$ & $-0.62   \pm   0.11 $ & $  -2.11  \pm  0.03  $ & $  -2.00 \pm 0.03  $ & $  -2.27 \pm 0.04  $ & $ -2.43 \pm0.10$\\ 
NGC~5024& M53   & $16.58\pm0.02$ & $-0.53   \pm   0.10 $ &  \dots                 &   \dots              & $  -2.06 \pm 0.09  $ & $ -2.02 \pm0.15$\\ 
NGC~5272& M3    & $15.44\pm0.03$ & $-0.42   \pm   0.10 $ &     \dots              &     \dots            & $  -1.50 \pm 0.05  $ & $ -1.50 \pm0.03$\\ 
NGC~5904& M5    & $14.98\pm0.03$ & $-0.23   \pm   0.10 $ & $  -1.38  \pm  0.05  $ & $  -1.12 \pm 0.03  $ & $  -1.33 \pm 0.02  $ & $ -1.26 \pm0.06$\\ 
NGC~6205& M13   & $14.73\pm0.05$ & $-0.35   \pm   0.11 $ & $  -1.63  \pm  0.04  $ & $  -1.33 \pm 0.05  $ & $  -1.58 \pm 0.04  $ & $ -1.60 \pm0.08$\\ 
NGC~6341& M92   & $14.62\pm0.02$ & $-0.60   \pm   0.10 $ &  \dots                 &   \dots              & $  -2.35 \pm 0.05  $ & $ -2.38 \pm0.07$\\      
NGC~6362&       & $15.52\pm0.02$ & $ +0.05  \pm   0.10 $ & $  -1.18  \pm  0.06  $ & $  -0.99 \pm 0.03  $ & $  -1.07 \pm 0.05  $ & $ -1.15 \pm0.10$\\
NGC~6723&       & $15.60\pm0.02$ & $-0.01   \pm   0.10 $ & $  -1.12  \pm  0.07  $ & $  -0.96 \pm 0.04  $ & $  -1.10 \pm 0.07  $ & $ -1.11 \pm0.10$\\ 
NGC~6752&       & $13.68\pm0.05$ & $-0.03   \pm   0.11 $ & $  -1.54  \pm  0.03  $ & $  -1.24 \pm  0.03  $ & $  -1.55 \pm 0.01  $ & $ -1.57\pm 0.10$\\ 
NGC~6809& M55   & $14.17\pm0.03$ & $-0.26   \pm   0.10 $ & $  -1.80  \pm  0.02  $ & $  -1.54 \pm 0.03  $ & $  -1.93 \pm 0.02  $ & $ -1.85 \pm0.10$\\ 
NGC~7089& M2    & $15.82\pm0.05$ & $-0.19   \pm   0.11 $ & $  -1.61  \pm  0.04  $ & $  -1.31 \pm 0.04  $ & $  -1.66 \pm 0.07  $ & $ -1.56 \pm0.10$\\ 
NGC~7099& M30   & $14.71\pm0.05$ & $-0.45   \pm   0.11 $ & $  -2.05  \pm  0.03  $ & $  -1.92 \pm 0.04  $ & $  -2.33 \pm 0.02  $ & $ -2.33 \pm0.10$\\ 
\enddata
\tablenotetext{a}{Iron abundances by Rutledge et al.\ (1997, RHS97) in the 
Zinn \& West (1984, ZW84) and in the Carretta \& Gratton (1997, CG97) metallicity scale.}  
\tablenotetext{b}{Iron abundances by Carretta et al (2009, C09).}  
\tablenotetext{c}{Iron abundances by Kraft \& Ivans (2003, 2004, KI04).}
\end{deluxetable}

%
%
%%%%%%%%%%%%%%%%%%%%%%%%%%%%%%%%%%%%%%%%%%%%%%%%%%%%%%%%%%%%%%%%%%%%%%%%%%%%%%%%%%%%%
% 			tab
%%%%%%%%%%%%%%%%%%%%%%%%%%%%%%%%%%%%%%%%%%%%%%%%%%%%%%%%%%%%%%%%%%%%%%%%%%%%%%%%%%%%%
\begin{deluxetable}{llccccc}
\scriptsize
\tablewidth{0pt}                       
\tablecaption{Evolutionary predictions for stellar structures constructed at fixed cluster 
age (12~Gyr), global metallicity and assuming different helium abundances.}
\tablehead{
\colhead{[M/H]}&
\colhead{$Z$}&
\colhead{$Y$}&
\colhead{$M_{pr}/M_{\odot}$\tablenotemark{a}}&
\colhead{$\Delta$ Y}\tablenotemark{b}}
\startdata
%  [M/H]    Z           Y     Mpr/Mo   Menv/Mo    Y_Iup    DeltaY
%#==================================================================
 $-1.79 $&$ 0.00030 $&$ 0.246  $&$0.8000   $& $0.009  $\\
 $-1.79 $&$ 0.00028 $&$ 0.300  $&$0.7300   $& $0.007  $\\
 $-1.79 $&$ 0.00026 $&$ 0.350  $&$0.6600   $& $0.005  $\\
 $-1.79 $&$ 0.00024 $&$ 0.400  $&$0.6000   $& $0.003  $\\

 $-0.96 $&$ 0.00200 $&$ 0.248  $&$ 0.8000  $& $0.014 $\\
 $-0.96 $&$0.00180  $&$0.300   $&$0.7600   $& $0.012  $\\
 $-0.96 $&$0.00170  $&$0.350   $&$0.7000   $& $0.009  $\\
 $-0.96 $&$0.00160  $&$0.400   $&$0.6000   $& $0.006  $\\

 $-0.35 $&$0.00800  $&$0.256   $&$0.9000   $& $0.020  $\\
 $-0.35 $&$0.00760  $&$0.300   $&$0.8000   $& $0.016  $\\
 $-0.35 $&$0.00700  $&$0.350   $&$0.7500   $& $0.013  $\\
 $-0.35 $&$0.06500  $&$0.400   $&$0.6700   $& $0.009  $\\

 \enddata
\tablenotetext{a}{Stellar mass of the progenitor at the tip of the RGB.}  
\tablenotetext{b}{Surface extra helium after the first dredge-up.}
\end{deluxetable}


\begin{thebibliography}{}


\bibitem[\protect\citeauthoryear{Alongi et al.}{1999}]{1a}Alongi, M., Bertelli, G., Bressan, A., Chiosi, C. 1991, A\&A, 244, 95
\bibitem[\protect\citeauthoryear{Alves \& Sarajedini}{1999}]{1b}Alves, D. R., Sarajedini, A. 1999, ApJ, 511, 225
\bibitem[\protect\citeauthoryear{Anderson} {2000}]{1c}Anderson, J. 2002, ASPC, 265, 87
\bibitem[\protect\citeauthoryear{Asplund et al.} {2005}]{1d}Asplund M., Grevesse N. \& Sauval A.J. 2005, in 
''Cosmic abundances as records of stellar evolution and nucleosynthesis'', 
eds. F.N. Bash \& T.J. Barnes, (San Francisco: ASP), 25
\bibitem[\protect\citeauthoryear{Bahcall et al.}{1995}]{1e}Bahcall J.N., Pinsonneault M.H. 1995, Rev. Mod. Phys 76,781
\bibitem[\protect\citeauthoryear{Bedin et al.} {2004}]{1f}Bedin, L. R., Piotto, G., Anderson, J. et al. 2004, ApJ, 605L, 125
\bibitem[\protect\citeauthoryear{Bergbush et al.} {2001}]{1g}Bergbusch, P. A., VandenBerg, D. A. 2001, ApJ, 556, 322   
\bibitem[\protect\citeauthoryear{Bono et al.} {2001}]{1h}Bono, G., Cassisi, S., Zoccali, M., Piotto, G. 2001, ApJ, 546L, 109 	   
\bibitem[\protect\citeauthoryear{Bono \& Castellani} {1992}]{1i}Bono, G., Castellani, V. 1992, A\&A, 258, 385
\bibitem[\protect\citeauthoryear{Briley et al.}{1996}]{1l}Briley, M. M., Suntzeff, N. B., Smith, V. V. et al. 1996, BAAS, 28, 1363
\bibitem[\protect\citeauthoryear{Buonanno et al.}{1986}]{1m}Buonanno, R., Caloi, V., Castellani, V. et al. 1986, A\&AS, 66, 79

\bibitem[\protect\citeauthoryear{Calamida et al.}{2009}]{1n} Calamida, A., Bono, G., Stetson, P. B, et al. 2009, ApJ, 706, 1277
\bibitem[\protect\citeauthoryear{Cariulo et al.}{2004}]{1o}Cariulo, P., Degl'Innocenti, S., Castellani, V. 2004, A\&A, 421, 1121 

%\bibitem[\protect\citeauthoryear{]{}Chaboyer, B. \& Krauss, L. M. 2002, ApJ, 567, L45 
\bibitem[\protect\citeauthoryear{Carretta \& Gratton}{1997}]{1p}Carretta, E., Gratton, R.G. 1997, A\&AS, 121, 95 (CG97)		
\bibitem[\protect\citeauthoryear{Carretta et al.}{2009}]{1q}Carretta, E., Bragaglia, A., Gratton, R. et al. 2009, 2009arXiv0910.0675 (C09) 
%'From Lithium to Uranium: Elemental Tracers of Early Cosmic Evolution', 
%eds. by V. Hill, P. François, F. Primas (Cambridge: Cambridge University Press), 389
\bibitem[\protect\citeauthoryear{Cassisi et al.}{1999}]{1r} Cassisi, S., Castellani, V., Degl'Innocenti, S., et al. 1999, A\&AS, 129, 267

\bibitem[\protect\citeauthoryear{Cassisi et al.}{1999}]{1s}Cassisi, S., Castellani, V., Degl'Innocenti, S., et al. 1999,A\&AS, 134, 103

\bibitem[\protect\citeauthoryear{Cassisi et al.}{2007}]{1t}Cassisi, S., Potekhin, A. Y., Pietrinferni, A. et al. 2007, ApJ, 661, 109
\bibitem[\protect\citeauthoryear{Cassisi \& Salaris}{1997}]{1u}Cassisi, S., Salaris, M. 1997, MNRAS, 285, 59
\bibitem[\protect\citeauthoryear{Cassisi et al.}{2002}]{1v}Cassisi, S., Salaris, M., Bono, G. 2002, ApJ, 565, 1231
\bibitem[\protect\citeauthoryear{Cassisi et al.}{2008}]{1z}Cassisi, S., Salaris, M., Pietrinferni, A. et al. 2008, ApJ, 672,L115
\bibitem[\protect\citeauthoryear{Castellani et al.}{1989}]{1w}Castellani, V., Chieffi, A., Norci, L. 1989,  A\&A, 216, 62
\bibitem[\protect\citeauthoryear{Castellani et al.}{1997}]{2a}Castellani, V., Ciacio, F., Degl'Innocenti, S., Fiorentini, G. 1997, A\&A, 322, 801
%\bibitem[\protect\citeauthoryear{]{}Charboyer, B. 2001, Pour la Science, 286, 60

\bibitem[\protect\citeauthoryear{Cohen \& Melandez}{2005}]{2b}Cohen, J., Melandez, J. 2005, AJ, 129, 303
\bibitem[\protect\citeauthoryear{Cottrell \& Da Costa}{1981}]{2c}Cottrell, P. L.,\& Da Costa, G. S. 1981, ApJ, 245L, 79
\bibitem[\protect\citeauthoryear{D'Antona et al.}{2005}]{2d}D'Antona, F., Bellazzini, M., Caloi et al. 2005, ApJ, 631, 868
\bibitem[\protect\citeauthoryear{D'Antona \& Caloi}{2008}]{2e}D'Antona,F., Caloi, V. 2008, MNRAS, 390, 693
\bibitem[\protect\citeauthoryear{Degl'Innocenti et al}{2008}]{deg08} Degl'Innocenti, S., Prada Moroni,
P.G., Marconi, M., Ruoppo, A. 2008, Ap\&SS, 316, 25
\bibitem[\protect\citeauthoryear{Denissenkov et al.}{2004}]{2f}Denissenkov, P. A., Weiss, A. 2004, ApJ, 603, 119
\bibitem[\protect\citeauthoryear{Desidera et al.}{1998}]{2g}Desidera, S., Bertelli, G., Ortolani, S. 1998,  IAU Symp. 189: 
'Fundamental Stellar Properties: The Interaction between Observation and Theory',
ed. by T.R. Bedding, Published by School of Physics, 164

\bibitem[\protect\citeauthoryear{Ferraro et al.}{1992}]{2h}Ferraro, F.R., Clementini, G., Fusi Pecci, F. et al. 1992, MNRAS, 256, 391
\bibitem[\protect\citeauthoryear{Ferraro et al.}{1999}]{2i}Ferraro, F.R., Messineo, M., Fusi Pecci, F. et al. 1999, AJ, 118, 1738 
\bibitem[\protect\citeauthoryear{Fusi Pecci et al.}{1990}]{Fusi}Fusi Pecci, F., Ferraro F. R., Crocker, D.A. et al. 1990, A\&A, 238, 95 (FP90)
\bibitem[\protect\citeauthoryear{Gratton et al.}{2004}]{2l}Gratton, R., Sneden, C., Carretta, E. 2004, ARA\&A, 42, 385
\bibitem[\protect\citeauthoryear{Grevesse \& Noels}{1993}]{2m}Grevesse, N. \& Noels, A. 1993, in ``Origin and Evolution of 
the elements'', ed. N. Prantzos, E. Vangioni-Flam, M. Cass\`e 
(Cambridge: Cambridge Univ. Press), 15
\bibitem[\protect\citeauthoryear{Guzik et al.}{2009}]{2n}Guzik, J. A., Keady, J.J., \& D.P. Kilcrease 2009, in Stellar Pulsation:
Challanges for Theory and Observation, ed. J.A. Guzik, P.A. Bradley (New
York: AIP), 577
\bibitem[\protect\citeauthoryear{Harris}{1996}]{2o}Harris, W. E. 1996, AJ, 112, 1487
\bibitem[\protect\citeauthoryear{Iben}{1968}]{2p}Iben, I. Jr. 1968, ApJ, 154, 581
\bibitem[\protect\citeauthoryear{Kraft \& Ivans}{2003}]{kraft03}Kraft, R.P., Ivans, I.I. 2003, PASP, 115, 143
\bibitem[\protect\citeauthoryear{Kraft \& Ivans}{2004}]{kraft04}Kraft, R.P., Ivans, I.I. 2004, arXiv:astro-ph/0305380v1
%\bibitem[\protect\citeauthoryear{]{}Kravtsov, V. V. 2009, AJ, 137, 5110
\bibitem[\protect\citeauthoryear{Landolt}{1992}]{land}Landolt, A. U., 1992, AJ, 104, 340
\bibitem[\protect\citeauthoryear{Marino et al.}{2008}]{2q}Marino, A. F.,Villanova, S., Piotto, G. et al. 2008, A\&A, 490, 625
\bibitem[\protect\citeauthoryear{Miloneet al.}{2008}]{2r}Milone, A. P., Bedin, L. R., Piotto, G. et al. 2008, ApJ, 673, 241
\bibitem[\protect\citeauthoryear{Norris}{2004}]{Norris}Norris, J.E. 2004, ApJ, 621, L57

%\bibitem[\protect\citeauthoryear{]{Lee}Pietrinferni, A., Cassisi, S., Salaris, M., Castelli, F. 2004, ApJ, 612, 168
\bibitem[\protect\citeauthoryear{Pietrinferni et al}{2006}]{Lee2}Pietrinferni, A., Cassisi, S., Salaris, M., Castelli, F. 2006, ApJ, 642, 797
\bibitem[\protect\citeauthoryear{Pietrinferni et al}{2009}]{Leer}Pietrinferni, A., Cassisi, S., Salaris, M. et al. 2009, ApJ, 697, 275
\bibitem[\protect\citeauthoryear{Piotto}{2008}]{Piotto08}Piotto, G. 2008, Mem. Soc. Astr. It., 79, 334
\bibitem[\protect\citeauthoryear{Piotto et al.}{2007}]{Piotto07}Piotto, G., Bedin, L. R., Anderson, J. et al. 2007, ApJ, 661, L53
\bibitem[\protect\citeauthoryear{Renzini \& Fusi Pecci}{1988}]{Ren 88}Renzini, A., Fusi Pecci, F. 1988, ARA\&A, 26, 199
\bibitem[\protect\citeauthoryear{Riello et al.}{2003}]{Riello}Riello, M., Cassisi, S., Piotto, G. et al. 2003, A\&A, 410, 553
\bibitem[\protect\citeauthoryear{Rutledge et al.}{1997}]{rut}Rutledge, G., Hesser, J., Stetson, P.B. 1997a, PASP, 109, 907

\bibitem[\protect\citeauthoryear{Salaris \& Cassisi}{2005}]{3a} Salaris, M., \& Cassisi, S.  2005, in 'Evolution of Stars and Stellar 
Populations', (New York: J. Wiley \& Sons ), 117   
 
\bibitem[\protect\citeauthoryear{Salaris et al.}{2008}]{3b} Salaris, M., Cassisi, S., \& Pietrinferni, A. 2008, ApJ, 678L, 25	
\bibitem[\protect\citeauthoryear{Salaris et al}{1993}]{3c} Salaris, M., Chieffi, A., Straniero, O. 1993, ApJ, 414 580
\bibitem[\protect\citeauthoryear{Salaris et al.}{2006}]{3d} Salaris, M., Weiss, A.,  Ferguson, J. W., Fusilier, D. J. 2006, ApJ, 645, 1131
\bibitem[\protect\citeauthoryear{Shetrone \& Keane}{2000}]{3e} Shetrone, M. D., \& Keane, M. J. 2000, AJ, 119, 840
\bibitem[\protect\citeauthoryear{Siegel et al.}{2007}]{3f} Siegel, M. H., Dotter, A., Majewski, S. R. et al. 2007, ApJ, 667, L57
\bibitem[\protect\citeauthoryear{Smith \& Wirth}{1991}]{3g} Smith, GH, \& Wirth GD 1991, PASP, 103, 75
\bibitem[\protect\citeauthoryear{Thomas}{1967}]{3h}Thomas, H. C. 1967, ZA, 67, 420
\bibitem[\protect\citeauthoryear{Thevenin \& Idiart}{1999}]{The} Thevenin, F., Idiart, T. P. 1999, ApJ, 521, 753

\bibitem[\protect\citeauthoryear{Vandenberg \& Bargbush}{2006}]{3i} VandenBerg, D. A., Bergbusch, P. A., \& Dowler, P. D.
2006, ApJS, 162, 375  
\bibitem[\protect\citeauthoryear{Ventura \& D'Antona}{2005}]{3l}Ventura, P., D'Antona, F. 2005, ApJ, 635, 149

\bibitem[\protect\citeauthoryear{Yong et al.}{2009}]{3m}Yong, D., Grundhal, F., D'Antona, F. et al. 2009, ApJ, 695, 62
\bibitem[\protect\citeauthoryear{Zinn \& West}{1984}]{ZW84}Zinn, R., West, M. J. 1984, ApJS, 55, 45 (ZW84) 
\bibitem[\protect\citeauthoryear{Zoccali et al.}{1999}]{Zinn 80}Zoccali, M., Cassisi, S., Piotto, G. et al. 1999, ApJ, 518, 49
\end{thebibliography}
\end{document}